\documentclass[aps,english,showpacs,11pt]{revtex4}
\usepackage[latin1]{inputenc}
\usepackage{ucs}
\usepackage{amsmath}
\usepackage{amsfonts}
\usepackage{amssymb}
\usepackage{graphicx}
\usepackage{picinpar}
\usepackage[]{hyperref}

\begin{document}
\title{Comparison between the numerical solutions and the Thomas-Fermi
  approximation for atomic-molecular Bose-Einstein condensates}

\author{L. S. F. Santos} 
\email{leosioufi@gmail.com} 
\affiliation{ Departamento de Ci\^{e}ncias Exatas e da Terra, Universidade
  Federal de S\~ao Paulo, Campus Diadema, Rua Prof. Artur Riedel, 275,
  Jd. Eldorado, 09972-270 Diadema, SP, Brazil}

\author{M. O. C. Pires}
\email{marcelo.pires@ufabc.edu.br}
\affiliation{Centro de Ci\^{e}ncias Naturais e Humanas, Universidade
  Federal do ABC, Rua Santa Ad\'elia 166, 09210-170, Santo Andr\'{e}, SP,
  Brazil}

\author{Davi Giugno} 
\email{dgiugno@if.usp.br}
\affiliation{Instituto de F\'{i}sica, Universidade de S\~ao Paulo, CP 66318,
  05315-970, S\~ao Paulo-SP, Brazil}

\begin{abstract}
We study the stationary solution of an atomic Bose-Einstein condensate
coupled coherently to a molecular condensate with both repulsive and
attractive interspecies interactions confined in an isotropic harmonic
trap. We use the Thomas-Fermi approximation and find four kinds of analytical solution for the cases. These analytical solutions are adopted as trial function for the diffusive numerical solution of the Gross-Pitaevskii equations. For the repulsive interspecies interaction, the case in which the atomic and molecular wavefunctions are out-phase, the densities have similar profiles for both methods, however, the case where the wavefunctions are in-phase, there are considerable difference between the density profiles. For the attractive interspecies interaction, there are two cases in the Thomas Fermi approximation where the wavefunctions are in-phase. One of them has numerical solution that agree with the approximation and the other does not have corresponding numerical solution.   
\end{abstract}


\maketitle

\section{Introduction}

The experimental observation of the Bose-Einstein condensate (BEC)
from ultra-cold and dilute alkali gases allowed a lot of important
applications for, and investigations on, cold atomic gases
\cite{and95,dav95,brad95}. Many researchers have intended to produce a
mixture of atomic and molecular BEC (AMBEC) with atom-to-molecule
conversion by the Raman photo-association process \cite{wyn00} or by
the Feshbach resonance technique \cite{pet98}. At present, it is
possible to convert fermionic and bosonic atoms into homonuclear
\cite{gre03} and heteronuclear \cite{web08,lu07} diatomic molecules
using the Feshbach resonance technique. Moreover, there is some
experimental support to the coherent binding of bosonic atoms into
molecules through a Raman photo-association process
\cite{wyn00}. Among these possibilities, only the molecular
BEC composed of fermionic atoms have been observed, i. e., the AMBEC
has not yet been detected in laboratory at the present moment.

One of the main reasons for the study of the AMBEC is the perspective
of understanding molecular formation and destruction stimulated by a
coherent coupling from macroscopic quantum states \cite{hei00}. Then,
a feature of the AMBEC is the relative quantum phase between atomic
and molecular wave-functions, which offers an atom-optic analogue to the
second-harmonic generation in non-linear optics \cite{cus01}. The
quantum relative phase appears in any mixture of coherent state BEC
with conversion between particles \cite{mat98,iso99}. Moreover, for
the references \cite{xia09,shi11}, another characteristic is that the AMBEC
presents the phase separation, where the AMBEC does not exist as a
mixture in equilibrium, but rather producing pure molecular BEC
and AMBEC domain structures.

In the mean-field model, the dynamics of the AMBEC is described by two
coupled Gross-Pitaevskii equations (GPE) for the macroscopic wave
function of the atomic and molecular species. Many authors use the
Thomas-Fermi approximation (TFA) for solving the GPE
\cite{xia09,cus01} in the limit where the number of particles is large. In
this approximation, the kinetic energy terms in the GPE are
neglected. The main advantage of this procedure is the possibility of
obtaining analytical solutions for the GPE.

In previous works, the validity of the TFA in the AMBEC was
investigated from different points of view.  In work \cite{cus01}, the
authors analysed an AMBEC coupled via photo-association process and
excluded the TFA for the self-trapped states where the atom-molecule
interaction is attractive and the relative phase is $\pi$. In
reference \cite{xia09}, the authors adopted the TFA in the GPE for the
AMBEC coupled via the Feshbach resonance. The authors concluded that
the TFA required a slow spatial variation of the AMBEC to be valid,
which fails to occur at the boundary of the two phases.  Then, the
previous works agree that there are problems in the spatial
distributions when one adopts the TFA. However, the reference
\cite{xia09} did not compare the numerical solution of the GPE for the
AMBEC to the TFA solutions at the same conditions. 

In this work, we have compared the TFA to the numerical solution of the GPE
for the AMBEC in the available conditions. 
This paper have been organized as follows. In section \ref{sec01}, we have presented
the model and we have derived the GPE. The TFA is shown in the section
\ref{sec02}. The numerical implementation and the discussion have been dealt
with in sections \ref{sec03} and \ref{sec04}, respectively. The
conclusion and remarks have been left for section \ref{sec05}.

\section{The model and the Gross-Pitaevskii equations}
\label{sec01}

We have considered that the AMBEC is within a spherical harmonic trap with
frequency $\omega$ for both species, where the atom-to-molecule
conversion can be either via Feshbach resonance or via Raman
photo-association. Moreover, we have accounted for the molecule mass as being twice
the atomic mass, $m$. In the mean-field theory, the total energy in
units of $\hbar\omega$ is given by:
\begin{eqnarray}
\langle H\rangle&=&\intop\left(-\frac{\zeta}{2}\phi_{1}^{*}(r)\nabla^2\phi_{1}(r)+\frac{1}{2}r^2|\phi_{1}(r)|^2+\frac{U_1}{2}|\phi_{1}(r)|^4\right)d^{3}r
+\nonumber
\\
&+&\intop\left(-\frac{\zeta}{4}\phi_{2}^{*}(r)\nabla^2\phi_{2}(r)+r^2|\phi_{2}(r)|^2+\frac{U_2}{2}|\phi_{2}(r)|^4\right)d^{3}r
+\nonumber\\
&+&\intop\left(U_{12}|\phi_{1}(r)|^{2}|\phi_{2}(r)|^{2}+\frac{\alpha}{2}\left(\phi_{1}^{2}(r)\phi_{2}^{*}(r)+\phi_{1}^{2*}(r)\phi_{2}(r)\right)\right)d^{3}r.
\label{eq:Hamiltonian}
\end{eqnarray}

Here $\phi_1(r)$ and $\phi_2(r)$ are the atomic and molecular wave
functions in units of $(\hbar/m\omega)^{-3/4}$. The spatial radial
coordinate $r$ is in units of $(m\omega/\hbar)^{1/2}$.  The parameter $\zeta$ can be either 0 or 1 depending on whether the TFA is used or not.  The parameters $U_{1}$, $U_{2}$ and $U_{12}$ are the
atomic, molecular and atomic-molecular interaction strengths,
respectively. We have considered that $U_{1}$ and $U_{2}$ are both positive
and $U_{12}$ can be either positive or negative.  We have defined the
parametric coupling strength $\alpha$ as a positive parameter because
the real term
$\left(\phi_{1}^{2}(r)\phi_{2}^{*}(r)+\phi_{1}^{2*}(r)\phi_{2}(r)\right)$
can be either positive or negative. The interaction strengths and the
parametric coupling strength are measured respectively in units of
$(\hbar\omega)^{-1}(\hbar/m\omega)^{-3/2}$ and
$(\hbar\omega)^{-1}(\hbar/m\omega)^{-3/4}$. We have neglected the energy for
creating a molecule from two atoms.

We have obtained the wave functions $\phi_{1}(r)$ and $\phi_{2}(r)$ from the variational method. We have used the linear functional $\langle H
\rangle-\mu \langle N\rangle$, where the chemical potential, $\mu$, is the Lagrange's multiplier and $\langle N\rangle=\langle N_{1}\rangle+2\langle
N_{2}\rangle =\intop(|\phi_{1}(r)|^{2}+2|\phi_{2}(r)|^{2})d^{3}r$ is the average of the total number of particles.
The variational method provides the following GPE,
\begin{eqnarray}
-\left(\frac{\zeta}{2r^{2}}\right)\frac{\partial}{\partial r}\left(r^{2}\frac{\partial\phi_{1}(r)}{\partial r}\right)+\left(-\widetilde{\mu}(r)+U_{1}|\phi_{1}(r)|^{2}+U_{12}|\phi_{2}(r)|^{2}\right)\phi_{1}(r)+\alpha\phi_{1}^{*}(r)\phi_{2}(r)=0,\label{eq:stationary 1}
\end{eqnarray}
\begin{eqnarray}
-\left(\frac{\zeta}{4r^{2}}\right)\frac{\partial}{\partial r}\left(r^{2}\frac{\partial\phi_{2}(r)}{\partial r}\right)+\left(-2\widetilde{\mu}(r)+U_{2}|\phi_{2}(r)|^{2}+U_{12}|\phi_{1}(r)|^{2}\right)\phi_{2}(r)+\frac{\alpha}{2}\phi_{1}^{2}(r)=0,\label{eq:stationary 2}
\end{eqnarray}
where $\widetilde{\mu}(r)=\mu-\frac{r^{2}}{2}$ is the effective local
chemical potential in units of $\hbar\omega$.

We have replaced $\phi_1(r) = |\phi_1(r)|e^{i \theta_1}$
and $\phi_2(r)=|\phi_2(r)|e^{i \theta_1}\beta$ in the equations
(\ref{eq:stationary 1}) and (\ref{eq:stationary 2}), where $\beta=e^{i\theta}$ is a function of the
relative phase, $\theta$, between $\phi_1(r)$ and $\phi_2(r)$. Then the two
equations read:
\begin{equation}
-\left(\frac{\zeta}{2r^{2}}\right)\frac{\partial}{\partial r}\left(r^{2}\frac{\partial|\phi_{1}(r)|}{\partial r}\right)+\left(-\widetilde{\mu}(r)+U_{1}|\phi_{1}(r)|^{2}+U_{12}|\phi_{2}(r)|^{2}\right)|\phi_{1}(r)|+\alpha\beta|\phi_{1}(r)||\phi_{2}(r)|=0,
\label{eq:stationary 1-1}
\end{equation}
\begin{equation}
-\left(\frac{\zeta}{4r^{2}}\right)\frac{\partial}{\partial r}\left(r^{2}\frac{\partial|\phi_{2}(r)|}{\partial r}\right)+\left(-2\widetilde{\mu}(r)+U_{2}|\phi_{2}(r)|^{2}+U_{12}|\phi_{1}(r)|^{2}\right)|\phi_{2}(r)|+\frac{\alpha}{2}\beta|\phi_{1}(r)|^{2}=0.\label{eq:stationary 2-1}
\end{equation}

The equations (\ref{eq:stationary 1-1}) and (\ref{eq:stationary 2-1}) can be solved only for $\beta=\pm1$. We have classified the solutions $\beta=1$ and $\beta=-1$ respectively as " in-phase" and "out-of-phase". The parametric coupling strength, $\alpha$ determines effectively the repulsion ($\beta=+1$) and the attraction ($\beta=-1$) between the atomic and molecular BEC. These results and classification have appeared in precedent work \cite{cus01}.

The solutions of the equations (\ref{eq:stationary 1-1}) and
(\ref{eq:stationary 2-1}) are different for each $\zeta$ value.  These
solutions have distinct behaviours and, therefore, should be analysed
separately. The analysis for each kind of solution shall be done in
the next two sections.

\section{Thomas-Fermi approximation }
\label{sec02}

Considering $\zeta=0$ in the equations (\ref{eq:stationary 1-1}) and
(\ref{eq:stationary 2-1}), we have found three kinds of solutions at a
specific position $r$. These kinds are: vacuum where
$|\phi_{1}(r)|=|\phi_{2}(r)|=0$ ($V$), pure molecular solution where
$|\phi_{2}(r)|^2=\frac{2\widetilde{\mu}(r)}{U_{2}}$ and
$|\phi_{1}(r)|=0$ ($PM$), mixed solution where
$|\phi_{1}(r)|\ne0$ and $|\phi_{2}(r)|\ne0$ ($AM$) . There is no  pure atomic solution where
$|\phi_{1}(r)|\ne0$ and $|\phi_{2}(r)|=0$ because $\alpha\ne0$. For
each radius $r$, the equations (\ref{eq:stationary 1-1}) and
(\ref{eq:stationary 2-1}) may have a different kind of solution.

The mixed solutions of (\ref{eq:stationary
  1-1}) and (\ref{eq:stationary 2-1}) can be found by solving two coupled
cubic equations \cite{xia09}.  For the sake of simplicity, we have restricted ourselves to the case where the atomic-molecular interaction strength is given by $U_{12}=\gamma\sqrt{U_{1}U_{2}}$,
where $\gamma=\pm1$. This restriction reduces the three solutions of
the cubic equations to just two. The situations
$\gamma=+1$ and $\gamma=-1$ correspond respectively to $U_{12}>0$
(atomic-molecular repulsion) and $U_{12}<0$ (atomic-molecular
attraction). Taking this into consideration, the mixed solutions are
given by,
\[
|\phi_{1}(r)|^{2}=\frac{\widetilde{\mu}(r)}{U_{1}}-\frac{\alpha\beta}{U_{1}}|\phi_{2}(r)|-\gamma\sqrt{\frac{U_{2}}{U_{1}}}|\phi_{2}(r)|^{2},
\]
\[
|\phi_{2}|=\beta\gamma B(r)+\delta\sqrt{B^{2}(r)+\gamma C(r)},
\]
where $\delta=\pm1$ labels the two solutions of (\ref{eq:stationary 1-1}) and (\ref{eq:stationary
  2-1}),
\[
B(r)=\frac{1}{3\alpha}\left(\widetilde{\mu}(r)\left(\gamma-2\sqrt{\frac{U_{1}}{U_{2}}}\right)-\frac{\alpha^{2}}{2\sqrt{U_{1}U_{2}}}\right),
\]
and
\[
C(r)=\frac{\widetilde{\mu}(r)}{3\sqrt{U_{1}U_{2}}}.
\]

We have classified the mixed solutions according to the sequence of signals
$AM_{\gamma \beta \delta}$. For example, the solution $AM_{-++}$ has
$\gamma=-1$, $\beta=+1$ and $\delta=+1$. In contrast with $\beta$ and
$\gamma$, the parameter $\delta$ does not indicate the
atomic-molecular attraction or repulsion. 

We have excluded those solutions that do not have a mathematical or physical
meaning. Firstly, on mathematical grounds we have excluded those solutions
for which $|\phi_{1}(r)|^{2}<0$, $|\phi_{2}(r)|<0$ and
$|\phi_{2}(r)|\notin\mathbb{R}$ ($B^2-\gamma C<0$). 
Next, we have rejected the solutions where the total density,
$|\phi_{1}|^{2}+2|\phi_{2}|^{2}$, increases with $r$. This is because the trap
forces the particles towards the center of the system. 

Another physical criterion is the continuity of the density.  To ensure
this property, it is necessary to define new continuous
functions $|\phi_{1}(r)|^{2}$ and $|\phi_{2}(r)|$, where a different
$r$ corresponds to a different solution. Indeed, we have one situation
where we are not able to build a continuous function for
$|\phi_{1}(r)|^{2}$ and $|\phi_{2}(r)|$.

Finally, we have ignored the trivial vacuum solution for all $r$ because it
does not satisfy the normalization condition $N\ne 0$.

According to these criteria, the solutions $AM_{++-}$ and
$AM_{---}$ are excluded for all $r$. The solutions $AM_{--+}$ and
$AM_{+--}$ are rejected by the mathematical criteria for
$r<\sqrt{2\mu}$ and by the physical criteria for $r>\sqrt{2\mu}$.

Then, we have five possible ways to
describe the atomic and molecular densities for all $r$ in the TFA when we have restricted $U_1>U_2/4$. This restriction is close to what is observed in realistic solutions. Otherwise, all the following analysis would be pointless. 
In the table \ref{tabela01} the classification of these five
possibilities is shown and the intervals of the different domain
solutions are indicated. The expressions for $\mu_\pm$ and $\mu_d$ that appeared in table \ref{tabela01} are defined by:  
\[
\mu_{d}=-\frac{\alpha^{2}}{\left(\sqrt{U_{2}}-2\sqrt{U_{1}}\right)^{2}}\left(\sqrt{\frac{U_{2}}{U_{1}}}+1-\sqrt{\frac{3}{4}\left(\frac{U_{2}}{U_{1}}\right)+3\sqrt{\frac{U_{2}}{U_{1}}}}\right),
\]
\[
\mu_{\pm}=\frac{2\alpha^{2}}{\left(\sqrt{U_{2}}\pm2\sqrt{U_{1}}\right)^{2}}.
\]

\begin{table}
\caption{Kinds of solution in the TFA.}
\begin{center}
\begin{small}
\begin{tabular}{|c|c|c|c|}
\hline 
{Classification}  &  $AM$ & $V$ & $PM$ \tabularnewline
\hline 
\hline
$S_M$ & - & $r\ge \sqrt{2\mu}$ & $r\le \sqrt{2\mu}$ \tabularnewline
\hline 
$S_{+++}$ & $r\le\sqrt{2\mu}$ & $r\ge\sqrt{2\mu}$ & -  \tabularnewline
\hline 
$S_{-+-}$ & $r\le\sqrt{2\mu}$ & $r\ge\sqrt{2\mu}$ & - \tabularnewline
\hline 
$S_{-++}$ for $\mu>\mu_+$ & $r\le\sqrt{2\left(\mu-\mu_{+}\right)}$ & $r\ge\sqrt{2\mu}$ & $\sqrt{2\left(\mu-\mu_{+}\right)}\le r\le\sqrt{2\mu}$ \tabularnewline
\hline 
$S_{+-+}$ &  $\sqrt{2\left(\mu-\mu_{-}\right)}\le r\le\sqrt{2\left(\mu-\mu_{d}\right)}$ & $r\ge\sqrt{2\left(\mu-\mu_{d}\right)}$ & $r\le\sqrt{2\left(\mu-\mu_{-}\right)}$ \tabularnewline
\hline 
\end{tabular}
\end{small}
\end{center}
\label{tabela01}
\end{table}

The wave functions $\phi_1(r)$ and $\phi_2(r)$, in the cases $S_M$, $S_{+++}$, $S_{-+-}$ and $S_{-++}$, are continuous. However the wave functions are discontinuous for the case $S_{+-+}$ at $r=\sqrt{2(\mu-\mu_d)}$.  
These discontinuities have already appeared in previous
works \cite{xia09}.

We have not represented the solution $S_M$ because this case is
similar to the single component BEC. For the other four possibilities,
we have compared the TFA to the numerical solution, and this is dealt
with in the next section.

\section{Numerical solution}
\label{sec03}

We have solved the equations (\ref{eq:stationary 1-1}) and
(\ref{eq:stationary 2-1}) for $\zeta=1$ by applying the relaxation
algorithm \cite{rup95,adh02,dal96,brt06} for the GPE. The algorithm
consists of a method that provides a numerical solution of the GPE by
considering an imaginary time variable, $\tau$. Redefining the
parametrized GPE by the imaginary time, we can obtain two coupled
non-linear diffusion equations. The propagation of the trial
functions, using this diffusion equation, provides the numerical
stationary solution at large imaginary times. In order to compare the
numerical solutions to the TFA solutions, we have used the
Thomas-Fermi profiles as the trial functions.

For the implementation of the relaxation algorithm, we have first
rewritten the GPE (\ref{eq:stationary 1-1}) and (\ref{eq:stationary
  2-1}) considering $U_{12}=\gamma\sqrt{U_1U_2}$, as
\begin{eqnarray}
\begin{cases} 
\frac{d^2\psi_1(r)}{dr^2}=-2\widetilde{\mu}(r)\psi_1(r)+2U_1
\left(\frac{\psi_1^3(r)}{r^2}\right)+2\gamma \sqrt{U_1U_2}
\left(\frac{\psi^2_2(r)\psi_1(r)}{r^2}\right)+2\beta \alpha
\left(\frac{\psi_2(r)\psi_1(r)}{r}\right)
\\ \frac{d^2\psi_2(r)}{dr^2}=-8\widetilde{\mu}(r)\psi_2(r)+4U_2
\left(\frac{\psi^3_2(r)}{r^2}\right)+4\gamma \sqrt{U_1U_2}
\left(\frac{\psi^2_1(r)\psi_2(r)}{r^2}\right)+2\beta \alpha
\left(\frac{\psi^2_1(r)}{r}\right)
\end{cases},
\label{gpe4}
\end{eqnarray}
where the functions $\psi_1(r)$ and $\psi_2(r)$ are related to the
atomic and molecular wave functions by $|\phi_1(r)|=\psi_1(r)/r$ and
$|\phi_1(r)|=\psi_1(r)/r$.

We have introduced the imaginary time variable in the equations
(\ref{gpe4}). This imaginary time is able to lead the trial function
to the stationary solution by a diffusion process. In fact, this
process must be possible if we consider the set of the non-linear
diffusion equations,
\begin{eqnarray}
\begin{cases} 
\frac{\partial \psi_1(r)}{\partial \tau}=\frac{\partial
  ^2\psi_1(r)}{\partial r^2}+2\widetilde{\mu}(r)\psi_1(r)-2U_1
\left(\frac{\psi_1^3(r)}{r^2}\right)-2\gamma \sqrt{U_1U_2}
\left(\frac{\psi^2_2(r)\psi_1(r)}{r^2}\right)-2\beta \alpha
\left(\frac{\psi_2(r)\psi_1(r)}{r}\right) \\ \frac{\partial
  \psi_2(r)}{\partial \tau} =\frac{\partial^2\psi_2(r)}{\partial
  r^2}+8\widetilde{\mu}(r)\psi_2(r)- 4U_2
\left(\frac{\psi^3_2(r)}{r^2}\right)-4\gamma \sqrt{U_1U_2}
\left(\frac{\psi^2_1(r)\psi_2(r)}{r^2}\right)-2\beta \alpha
\left(\frac{\psi^2_1(r)}{r}\right)
\end{cases}.
\label{difusion}
\end{eqnarray}   
These equations provide a convergent diffusion process and several
solutions coexist within numerical precision for a large imaginary
time $\tau\to \infty$. At this limit, the trial solutions propagate to
the solution of equations (\ref{eq:stationary 1-1}) and
(\ref{eq:stationary 2-1}).

For these new functions $\psi_1$ and $\psi_2$, the normalization
conditions are given by,
\begin{eqnarray}
\begin{cases}
\int_0^\infty\psi^2_1(r)dr=\frac{N_1}{4\pi}
\\ \int_0^\infty\psi^2_2(r)dr=\frac{N_2}{4\pi}
\end{cases},
\end{eqnarray}
with the requirement that the wave functions must vanish far from the
trap center. In the same way, the non-linear term inside
Eq. (\ref{gpe4}) must eventually become negligible compared to the
other two terms. The asymptotic form has the behaviour
$\psi_1(r)\approx c_1e^{-r^2/4+(\mu -1/2)\ln(r)}$ and
$\psi_2(r)\approx c_2e^{-r^2/4+(4\mu-1/2)\ln(r)}$, where
$c_1=\sqrt{N_1/\Gamma(\mu)}$ and $c_2=\sqrt{N_2/(2\Gamma(4\mu))}$ were
determined by the wave function normalization, with $\Gamma(x)$ being
the gamma function.

For the limit $r\to 0$, the non-linear term inside Eq. (\ref{gpe4}) approaches a
constant due to the regularity of the wave function at $r=0$. Then we
can write $\psi_1(r)\approx \psi_1'(0)r$ and $\psi_2(r)\approx
\psi_2'(0)r$ in this limit.

We have kept the total particle number fixed in the whole diffusion
process. However, the chemical potential is changed at every iteration
step. The convergence of the chemical potential has been used as a
criterion to stop the diffusion process.

\section{Discussion}
\label{sec04}

We have compared the TFA to the numerical solution for each spatial
distribution. This correspondence of the TFA to the numerical solution is given by the input of the diffusive numerical procedure. 

The numbers of particles in our simulation are compatible with the usual numbers of particles of experimental atomic BEC's, namely, $10^4$, $2\times10^4$, $5\times 10^4$, $10^5$, $2\times10^5$, $5\times 10^5$ and $10^6$. We have adjusted the chemical potential in order to obtain these numbers of particles.

For the repulsive atom-molecule interaction ($U_{12}>0$) corresponding to $\gamma=+1$  cases, we have employed
the parameters used in theoretical work \cite{xia09} which is based on
values in \cite{hei00}, $U_1= 0.1$, $U_2 = 0.036$ and $\alpha = 1.08$. For the attractive atom-molecule interaction ($U_{12}<0$) corresponding to $\gamma=-1$, we have used those from \cite{wyn00,cus01} instead,
namely $U_1 = 0.062$, $U_2 = 0.12$ and $\alpha=1.09$. As for the
Cusack data \cite{cus01}, the only difference between our set of
parameters and theirs is the value of $U_{12}$. We have been set to
$U_{12} =- \sqrt{U_1 U_2}\approx-0.0876$ but the reference
\cite{cus01} had been fixed exactly at $U_{12}=-0.087$. Even
so, our value for $U_{12}$ is compatible with the experimental
measurement $U_{12}=-0.087\pm0.07$ \cite{wyn00,cus01}. All sets of
parameters are inspired on the diatomic molecules
created at rest in a dilute Bose-Einstein condensate of rubidium-87
atoms with coherent free-bound stimulated Raman transitions
\cite{wyn00,hei00}.

\begin{figure}[!htp]
\begin{center}

\includegraphics[width=12cm]{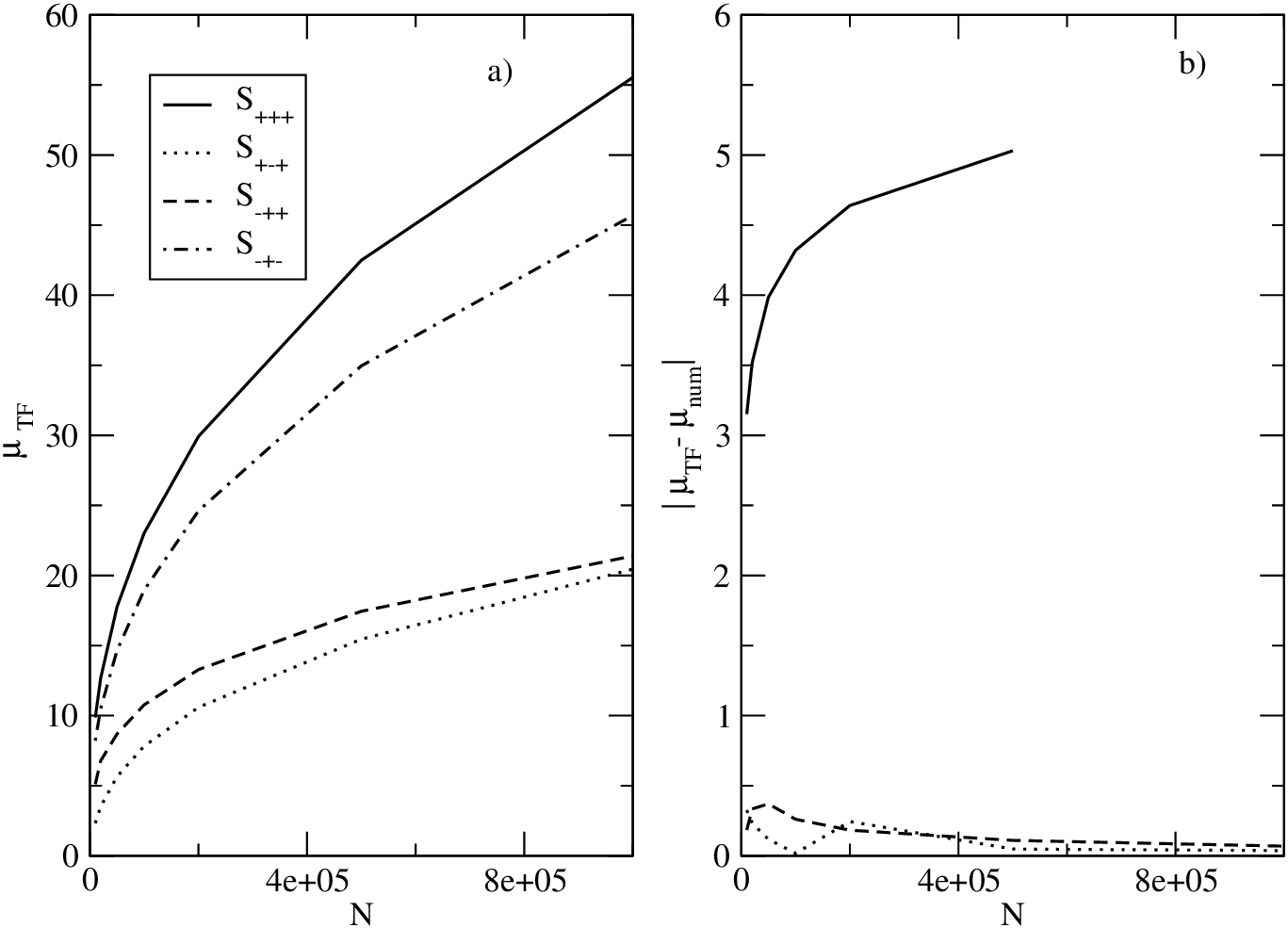}
\caption{The chemical potential in the TFA (fig. a)) and its variation between the TFA and numerical solution (fig. b)), $|\mu_{TF}-\mu_{num}|$ as functions of the number of particles in four cases, $S_{+++}$ (solid line), $S_{+-+}$ (dotted line), $S_{-++}$ (dashed line) and $S_{-+-}$ (dotted-dashed line). There is no variation for the $S_{-+-}$ case because we do not have the numerical solution for this case.}
\label{mu}

\end{center}
\end{figure}

\begin{figure}[!htp]
\begin{center}

\includegraphics[width=12cm]{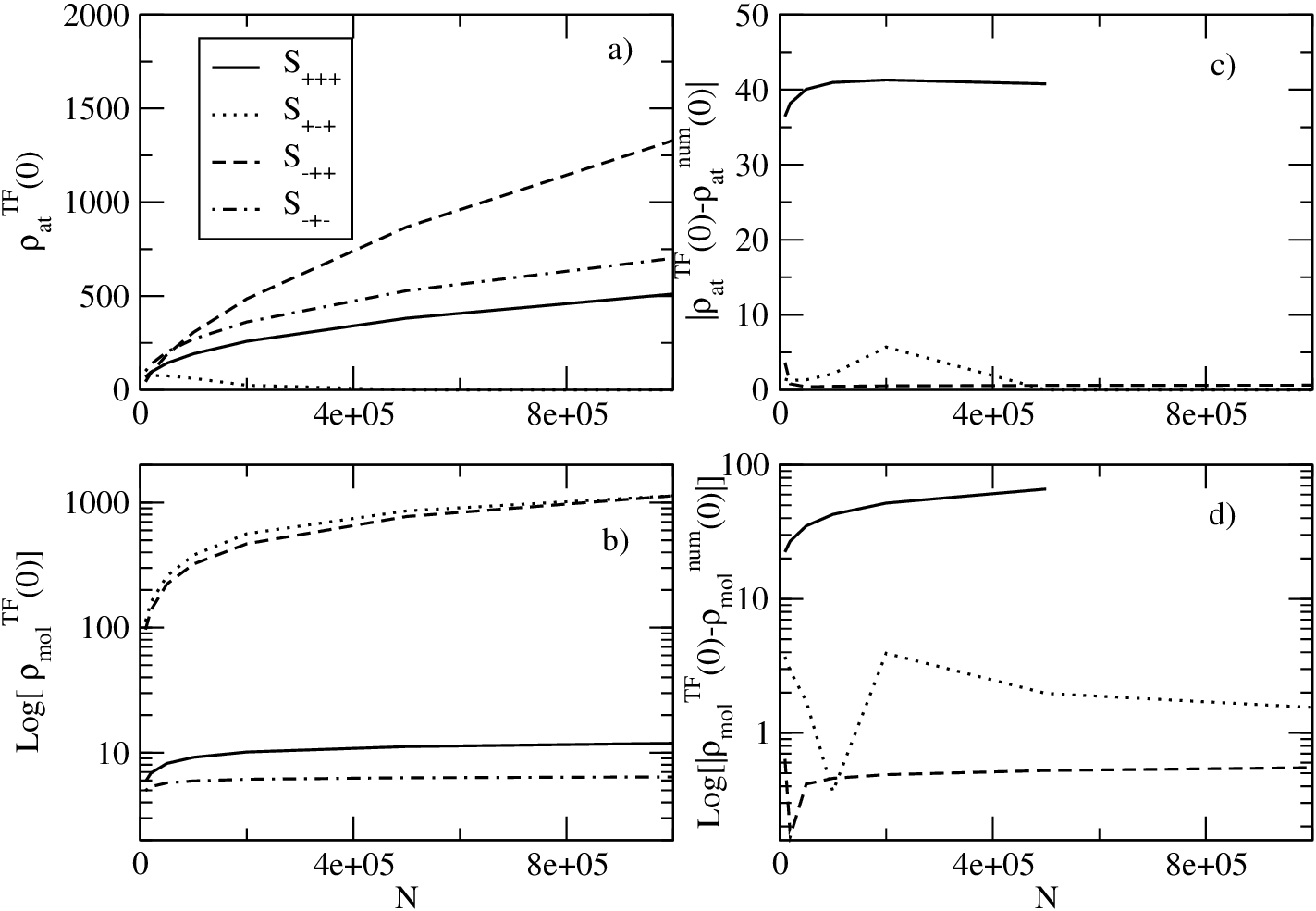}
\caption{The atomic (fig. a)) and molecular (fig.b)) central density in the TFA and the modular variation of atomic (fig. c)) and molecular (fig. d)) central densities, $|\rho^{at}_{num}(0)-\rho^{at}_{TF}(0)|$ and $|\rho^{mol}_{num}(0)-\rho^{mol}_{TF}(0)|$ as functions of the number of particles in four cases, $S_{+++}$ (solid line), $S_{+-+}$ (dotted line), $S_{-++}$ (dashed line) and $S_{-+-}$ (dotted-dashed line). There is no variation in the $S_{-+-}$ because we do not have numerical solution for this case. We have not found an available numerical solution for $S_{+++}$ at $N>7\times 10^5$.}
\label{dens}

\end{center}
\end{figure}

For these parameters, the behaviour of the chemical potential and the central density as function of number of particles can be seen in figures \ref{mu} and \ref{dens}. In the figure \ref{mu}, we have graphed the chemical potential in the TFA and the modular variation of the chemical potential obtained by the TFA solution and by the diffusive numerical solution. In the TFA, the chemical potential of the $S_{+++}$ and $S_{-+-}$ cases is larger than  $S_{-++}$ and $S_{+-+}$ cases for any numbers of particles. Analysing the figure \ref{mu}b), the $S_{+++}$ case present the modular variation more significant than $S_{-++}$ and $S_{+-+}$ cases.  Notice that for the $S_{-+-}$ we have no numerical solutions at all. Hence we shall not display any relative variation curves.

Furthermore, in the fig. \ref{dens}, we have produced the graphs for the atomic and molecular central density as function of the number of particles in the TFA. In graphs \ref{dens}a) and \ref{dens}b), four cases are considered. In the graphs \ref{dens}c) and \ref{dens}d), we have plotted the modular variation of the atomic and molecular central density.  In both, the modular variation of the central density in the $S_{+++}$ case is larger than the others.   

\vspace{0.5cm}

\vspace{0.5cm}

\begin{figure}[!htp]
\begin{center}

\includegraphics[width=12cm]{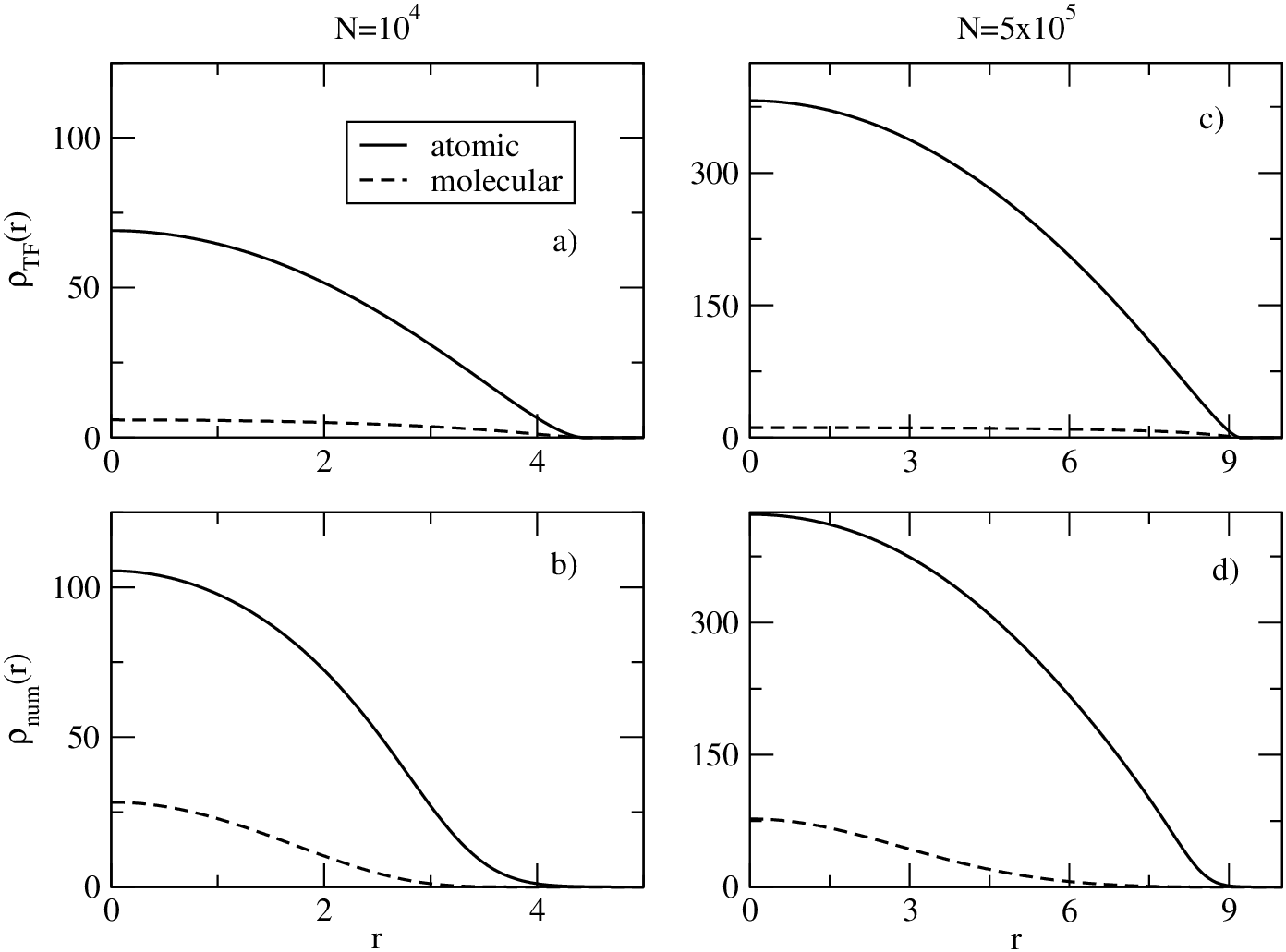}
\caption{Atomic density plot (solid line) and molecular density plot (dashed line) of the TFA and the numerical solution for $S_{+++}$ considering $10^{4}$ particles (graphs a) and b)) and $5\times 10^{5}$ particles (graphs c) and d)). We have chosen the initial parameters $U_1=0.1$, $U_2=0.036$ and $\alpha=1.08$.}
\label{s+}

\end{center}
\end{figure}

For the $S_{+++}$ case, there is not available numerical solution for $N>7\times 10^5$ ($\mu>47.$). In this case, there is a considerable difference in the
central densities of the profiles provided by the numerical
calculation and the TFA (see in the fig. \ref{dens} and \ref{s+}). The diffusion process in the numerical solution concentrate the
densities in the center of the trap. Moreover, in the TFA, the
molecular and atomic densities are found to be continuous, but their
first spatial derivatives present a first-order discontinuity when
$r=\sqrt{2\mu}$. On the other hand, we have found
that the densities and the first derivatives are continuous everywhere
in the numerical solution.

\vspace{1.0cm}

\begin{figure}[!htp]
\begin{center}

\includegraphics[width=12cm]{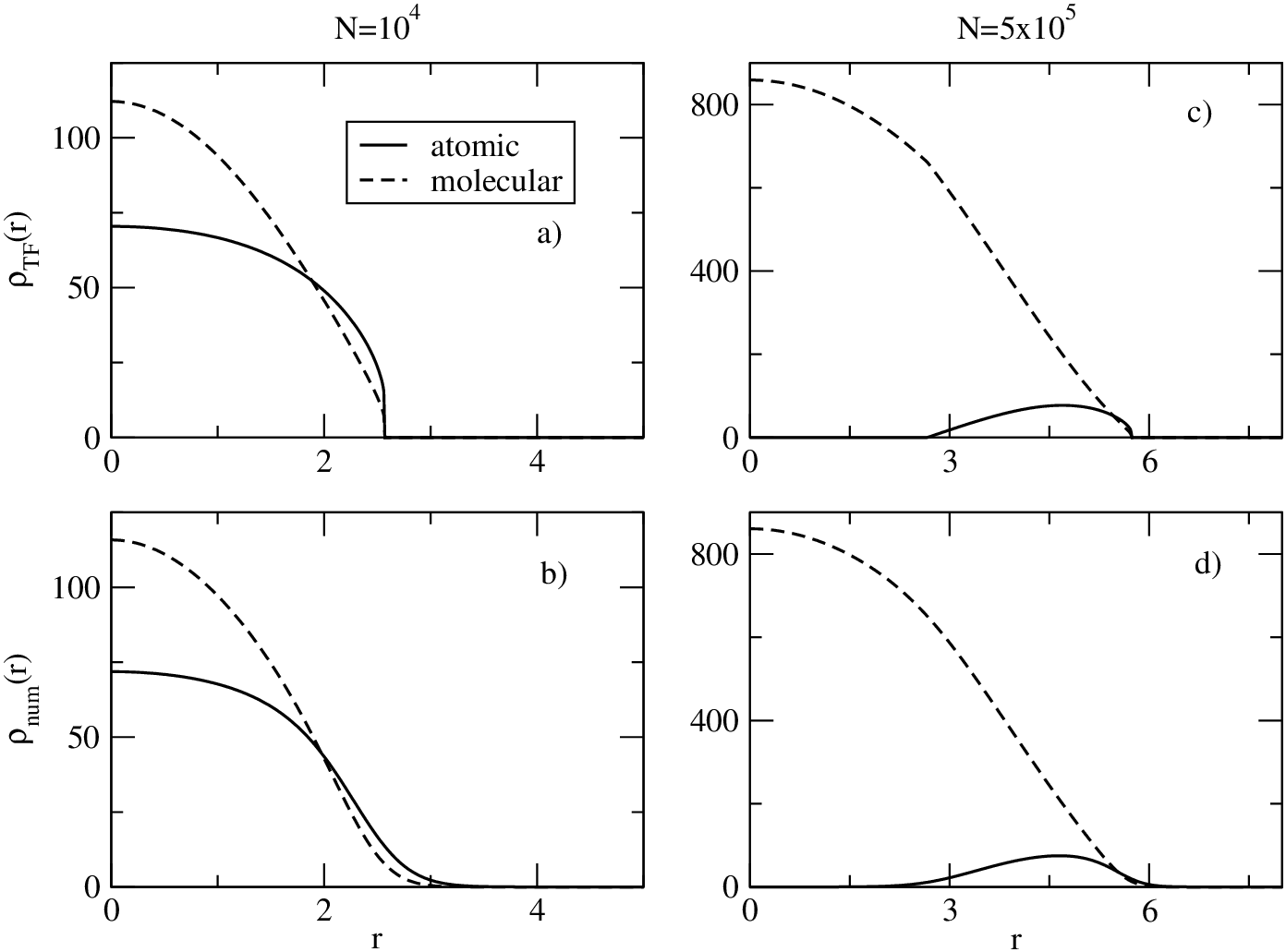}
\caption{Atomic density plot (solid line) and molecular density plot
  (dashed line) for $S_{+-+}$ considering $10^{4}$ particles (graphs a) and b)) and $5\times 10^{5}$ particles (graphs c) and d)). We have chosen the initial parameters $U_1=0.1$, $U_2=0.036$ and $\alpha=1.08$.}
\label{s+-}

\end{center}
\end{figure}

For the $S_{+-+}$ case (see in the fig. \ref{s+-})
we have found a good agreement between the TFA and the numerical
solution. Although the agreement, for the TFA there are discontinuities not only
in the first spatial derivative of the spatial distributions but also
in the distributions themselves at the point
$r_{d}=\sqrt{2(\mu-\mu_{d})}$ and, for $r>r_{d}$, there is
vacuum. Moreover, in the case where $\mu>\mu_-$ (for our parameters, $\mu>10.90$), the $S_{+-+}$ case at the TFA presents a
discontinuity in the first spatial derivatives of both the atomic and
molecular densities, at the point $r_{-}=\sqrt{2(\mu-\mu_{-})}$. For
$r<r_{-}$, the spatial distribution becomes purely molecular. While in  the
numerical solution, there is no phase separation between the atomic
and molecular distribution, and no discontinuities of any kind have been
seen. However, in the region of purely molecular density in the TFA,
the atomic density provided by the numerical solution is almost null.

\vspace{1.0cm}

\begin{figure}[!htp]
\begin{center}

\includegraphics[width=12cm]{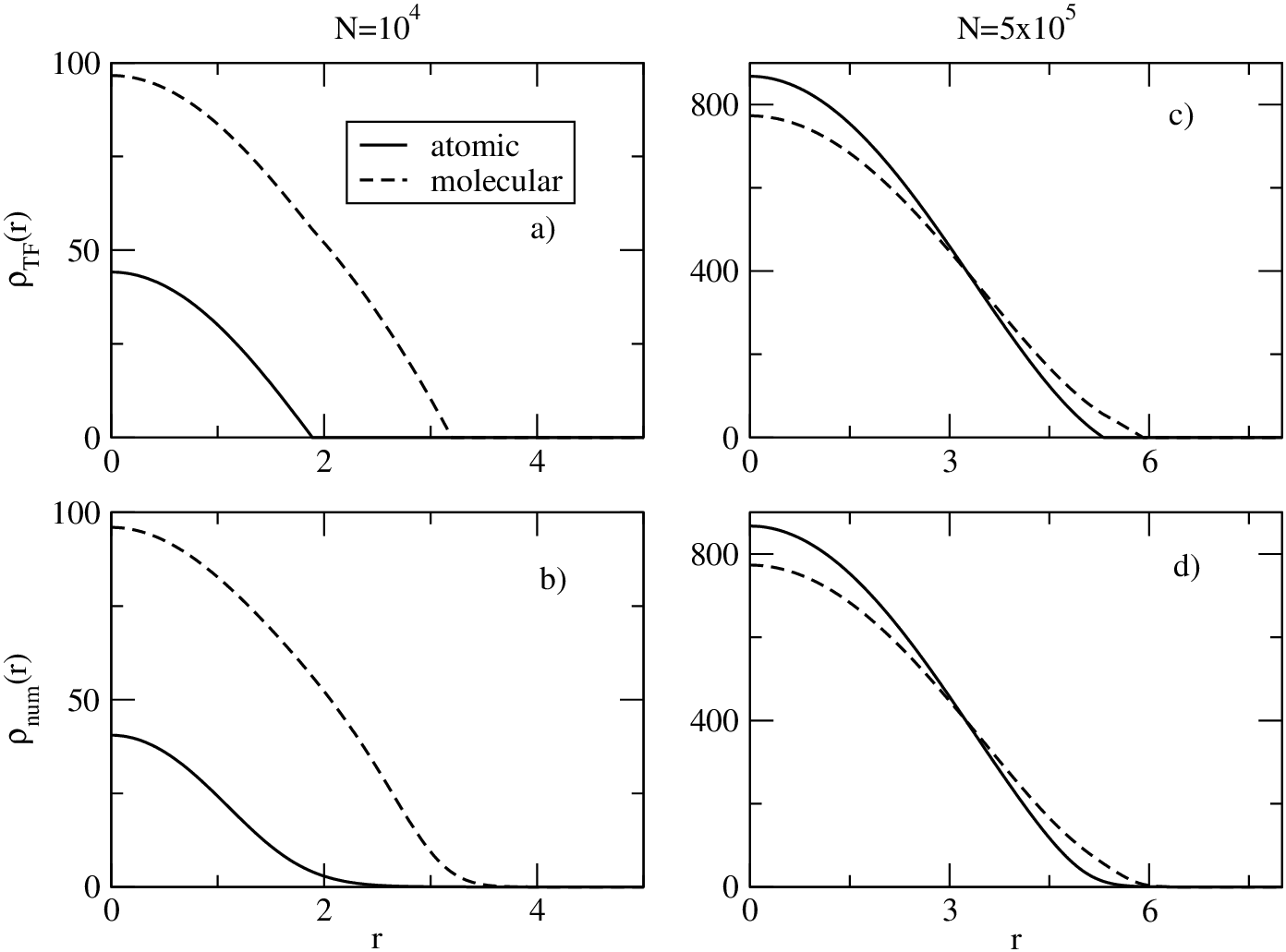}
\caption{Atomic density plot (solid line) and molecular density plot
  (dashed line) for $S_{-++}$ considering $10^{4}$ particles (graphs a) and b)) and $5\times 10^{5}$ particles (graphs c) and d)). We have chosen the initial parameters $U_1=0.062$, $U_2=0.012$ and $\alpha=1.09$.}
\label{s+--M}

\end{center}
\end{figure}

Likewise the $S_{+-+}$, the $S_{-++}$ case in the TFA
(see in the fig. \ref{s+--M}) has good agreement to the corresponding numerical solution. Although there are first-derivative discontinuities at two distinct
points for the TFA, namely at $r_{+}=\sqrt{2(\mu-\mu_{+})}$ and at
$r_{1}=\sqrt{2\mu}$. For $r<r_{+}$, the solution is mixed
(atomic-molecular); for $r_{+}<r<r_{1}$, it is purely molecular. For
$r>r_{1}$, there is a vacuum. Analogously to the $S_{+-+}$ case with $\mu>\mu_-$, the
numerical solution for the $S_{-++}$ case is made up of an
atom-molecule mixture. In the region of purely molecular density at
the TFA, the atomic density provided by the numerical solution is
almost null.

\vspace{0.5cm}

\begin{figure}[!htp]
\begin{center}

\includegraphics[width=12cm]{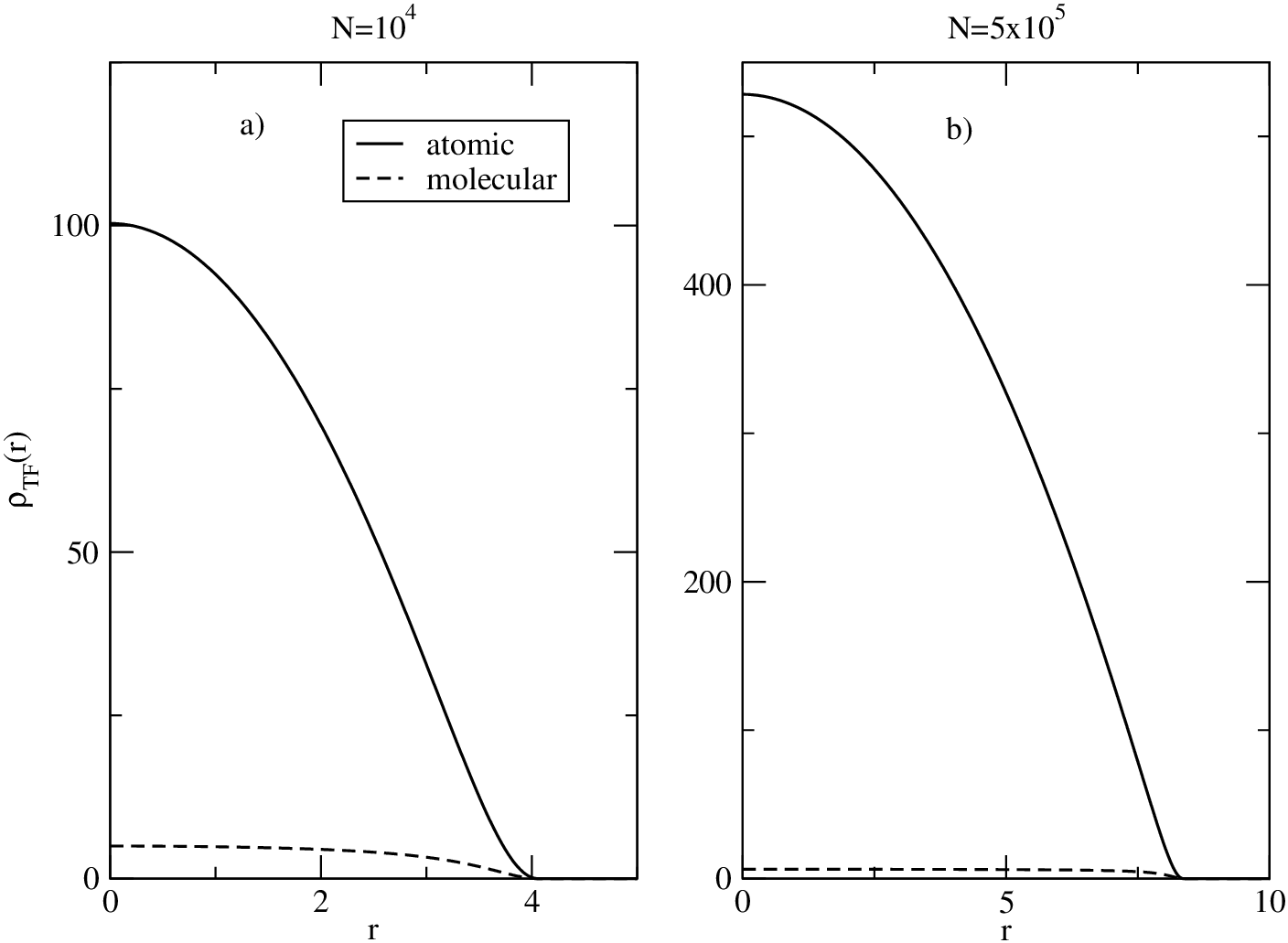}
\caption{Atomic density plot (solid) and molecular density plot
  (dashed) for $S_{-+-}$ considering $10^{4}$ particles (graphs a)) and $5\times 10^{5}$ particles (graphs b)) in the TFA. We have chosen the initial parameters $U_1=0.1$, $U_2=0.036$ and $\alpha=1.08$. There is no numerical solution for this case.}
\label{s+--}

\end{center}
\end{figure}

For the case $S_{-+-}$, (see in Fig. \ref{s+--}) there are discontinuities in the first
derivatives of the densities in the TFA, as seen in the other solutions.
Differently from the other cases, there is no available numerical
solution when we used the TFA as a trial function for the parameters
in the references \cite{cus01,xia09}.

\vspace{0.5cm}

\section{Conclusion}
\label{sec05}

We have considered a Bose gas in the AMBEC. The system consisting of atoms coherently coupled with their
homonuclear diatomic molecules at zero temperature. For the TFA, we have analysed four
possible spatial distributions in the conditions where the
interspecies interactions are either negative or positive and where
the coupling mode can be either in-phase or out-phase. For each
solution at the TFA, we have used as trial function to obtain the numerical solution for the GPE.

As seen in the reference \cite{xia09}, the TFA spatial distribution
was that it admits discontinuities in the derivatives of the densities
in every case and presents discontinuities in the densities
themselves in the  $S_{+-+}$ case. There are phase
separations in every spatial distributions: a mixture-purely molecular
interface in the cases $S_{+-+}$ with $\mu=\mu_-$ and $S_{-++}$, a vacuum-mixture
interface in every case except for $S_{-++}$, where there is a
vacuum-purely molecular separation.

The TFA is suitable with the numerical solution for the $S_{+-+}$ and $S_{-++}$ cases that present the phase separation.
On the contrary of the solution at the TFA, the numerical solutions
did not become zero neither for the molecular nor the atomic
densities. Consequently, there was no phase separations of any
kind. Despite this difference, in the regions where the TFA has the
purely molecular distribution, the numerical atomic density
contributed with a low fraction of the total density compared to the
molecular density. And, in the case where there is vacuum in the TFA,
the numerical solution presents a low total density. 

We have found the numerical solution corresponding to the $S_{+++}$ case for $\mu<47.$. For these numerical solutions, the atomic and molecular density profiles present a considerable difference between the TFA profiles. We have noted the atomic and molecular concentration in the center of the trap provided by the numerical solution is higher than the concentration generated by the TFA solution. This fact is unusual because we hope that the TFA solution will converge to the numerical solution when the number of particle increases. Likewise the cases above, the numerical solution has low total density in the region where the TFA presents vacuum.

In the condition of $\gamma=-1$ and $\beta=+1$ we have found two available solution in the TFA, $S_{-+-}$ and $S_{-++}$. For the $S_{-+-}$ case, there is not numerical solution. As well as, the reference \cite{cus01} have found only one numerical solution for this specific external condition. Thus we have related the $am_+$ state determining by the reference \cite{cus01} with the $S_{-++}$ case.  

We have noticed for the same kind of atom-molecule interaction ($\gamma$), the solution with the less chemical potential have accordance between the TFA and the numerical solutions. For $\gamma=+1$, the atomic and molecular densities in the TFA and the numerical solution are agreed to each other only for $S_{+-+}$, while it does not happen in the $S_{+++}$ cases. In the attractive atom-molecule interaction, $\gamma=-1$, the TFA solution of the $S_{-++}$ case has accorded to the numerical solution. The $S_{-+-}$ not even have found the numerical solution.     

For the relative phase equal to $\pi$ ($\beta=-1$) and the atom molecule interaction being attractive ($\gamma=-1$), we could not obtain solutions for the
TFA because $AM_{---}$ and $AM_{--+}$ both violated the physical and
mathematical criteria for the acceptability of a solution. This
situation corresponds to $am_-^t$ state of the reference
\cite{cus01} have concluded that it cannot be determined by the TFA.

We believe that our work clarifies the importance of being judicious
in the usage of the TFA. For the  
$S_{+-+}$ and $S_{-++}$ cases, the TFA is suitable to  describe the system. However, though we have found the numerical solution only for $S_{+++}$ case, the TFA can not describe the $S_{+++}$ and $S_{-+-}$ cases.

\begin{acknowledgments}
The authors would like to thank A. Gammal for tips on computing and
A. F. R. de Toledo Piza for helping in the early phase of the present
work. L. S. F. S. thanks FAPESP for financial support.
\end{acknowledgments}

\end{document}